\documentclass[manuscript, authorversion]{acmart}

\usepackage{graphicx}
\usepackage{subfig}
\usepackage{multirow}
\AtBeginDocument{%
  \providecommand\BibTeX{{%
    \normalfont B\kern-0.5em{\scshape i\kern-0.25em b}\kern-0.8em\TeX}}}
\newcommand{\xhdr}[1]{\vspace{1mm}\noindent{{\bf #1.}}}

\acmConference[Websci '24]{ACM Web Science Conference}{May 21--24, 2024}{Stuttgart, Germany}
\acmBooktitle{ACM Web Science Conference (Websci '24), May 21--24, 2024, Stuttgart, Germany}\acmDOI{10.1145/3614419.3644012}
\acmISBN{979-8-4007-0334-8/24/05}

\usepackage{todonotes}

\begin{document}

\title{A Longitudinal Study of Italian and French Reddit Conversations Around the Russian Invasion of Ukraine }

\author{Francesco Corso}
\affiliation{%
  \institution{Politecnico di Milano}
  \city{Milan}
  \country{Italy}}
\email{francesco.corso@polimi.it}

\author{Giuseppe Russo}
\affiliation{%
  \institution{Zurich ETH}
  \city{Zurich}
  \country{Switzerland}
}\email{russog@ethz.ch.}

\author{Francesco Pierri}
\affiliation{%
  \institution{Politecnico di Milano}
  \city{Milan}
  \country{Italy}
}\email{francesco.pierri@polimi.it}

\renewcommand{\shortauthors}{Corso, Russo, Pierri}

\begin{abstract}

Global events like wars and pandemics can intensify online discussions, fostering information sharing and connection among individuals. However, the divisive nature of such events may lead to polarization within online communities, shaping the dynamics of online interactions. Our study delves into the conversations within the largest Italian and French Reddit communities, specifically examining how the Russian invasion of Ukraine affected online interactions. We use a dataset with over 3 million posts (i.e., comments and submissions) to (1)  describe the patterns of moderation activity and
(2) characterize war-related discussions in the subreddits. We found changes in moderators' behavior, who became more active during the first month of the war. Moreover, we identified a connection between the daily sentiment of comments and the prevalence of war-related discussions. These discussions were not only more negative and toxic compared to non-war-related ones but also did not involve a specific demographic group.
Our research reveals that there is no tendency for users with similar characteristics to interact more. Overall, our study reveals how the war in Ukraine had a negative influence on daily conversations in the analyzed communities. This sheds light on how users responded to this significant event, providing insights into the dynamics of online discussions during events of global relevance.

\end{abstract}

\ccsdesc[500]{Information systems~Social networks}

\keywords{Reddit, sentiment analysis, Russian invasion of Ukraine, online moderation}

\maketitle
\noindent\textcolor{red}{Please remember to cite the published version of this paper:}

\textbf{\textcolor{red}{"Francesco Corso, Giuseppe Russo, Francesco Pierri" A Longitudinal Study of Italian and French Reddit Conversations Around the Russian Invasion of Ukraine, Proceedings of the 16th ACM Web Science Conference 2024"}}
\section{Introduction}

On February 24th, 2022, the Russian army invaded Ukraine, after years of tension and conflict that began with the annexation of Crimea in 2014. Since then, the war in Ukraine has caused over 10,000 casualties \footnote{https://news.un.org/en/story/2023/11/1143852 - Accessed on 29/11/2023}and forced over six million Ukrainians to abandon their country.
This action drew widespread condemnation from the international community, as many countries viewed it as an infringement on Ukraine's sovereignty and territorial integrity \footnote{https://www.consilium.europa.eu/en/policies/eu-response-ukraine-invasion/\#invasion - Accessed on 6/12/2023}.
Concurrently with the invasion, Russia became actively involved in spreading propaganda and misinformation regarding the conflict, aiming to manipulate public sentiment and erode support for Ukraine \cite{geissler2023russian}.  Such online information disorders and coordinated harm on social platforms are particularly significant during times of crisis when access to accurate and reliable information is essential \cite{pierri2020diffusion, di2022vaccineu}.
 
Due to the gravity of the conflict and the potential impact on regional stability, European countries, such as Italy and France, collectively pledged diplomatic and military support to Ukraine. This united front aimed not only to condemn the breach of Ukraine's sovereignty but also to deter an extension of the conflict to other European countries. 
This decision sparked a heated debate about the stance European countries should take regarding the Ukrainian war, quickly spreading to social media platforms like Twitter and Reddit, where most users took sides on either supporting military intervention or advocating for more diplomatic solutions \cite{ proedrou2010ukraine}. 
Therefore, examining the various stances taken by users in online debates can enhance our understanding of public opinion on this controversial issue.

\xhdr{Present Work} We carry out a longitudinal study of conversations around the war taking place on Reddit, which is a discussion platform that allows users to engage in discussions and share content \footnote{https://thebrandhopper.com/2023/03/16/the-rise-of-reddit-how-the-platform-became-cultural-phenomenon/ - Accessed on 23/08/2023} in various communities known as "subreddits."
Specifically, we analyze the subreddits \texttt{r/italy} and \texttt{r/france} as these two countries are amongst those states that more strongly support the Ukrainian defense against the Russian invasion\footnote{https://www.aljazeera.com/news/2023/2/16/mapping-where-every-country-stands-on-the-russia-ukraine-war - Accessed on 6/12/2023}. 
In both countries, the debate around the support of Ukraine against Russia was heated, reflecting the broader European sentiment towards the conflict. Our study analyzes submissions and comments in these subreddits to understand how the war affected users' behavior in these two subreddits. 
We articulate our contributions into three research questions.

\begin{itemize}
    \item \textbf{RQ1}: How does moderators' activity change after the start of the invasion? 
    \item \textbf{RQ2}: How does the invasion of Ukraine affect interactions between Reddit users?
    \item \textbf{RQ3}: Do we find evidence of homophily among Reddit users discussing the invasion?
\end{itemize}

To answer these questions, we collected 1.1M comments and 2.2M comments from \texttt{r/italy} and \texttt{r/france}, respectively. 
First (\textbf{RQ1}), to understand moderators' activity after the start of the war, we investigate the prevalence of moderated comments during 2022. We find a positive correlation in \texttt{r/italy}  between the number of comments discussing the war and those removed, with a peak of over 10\% of moderated comments on the same day of the invasion, while in \texttt{r/france} this result does not apply, since there is not a significant correlation.

Second \textbf{(RQ2)}, we investigate the interactions between users by analyzing the text of comments posted in each subreddit. We employ state-of-the-art multilingual BERT-based models to infer the sentiment and toxicity of comments, finding that those discussing the war are significantly more negative and more toxic than others. Moreover, we find an overall negative sentiment with a significant negative peak value at the beginning of the invasion.

Third (\textbf{RQ3}), drawing upon results from the previous research question, we build and analyze the network of interactions between users. We assign an average sentiment score to each user based on their comments. Additionally, we consider three social scores\cite{Waller_2021} (age, gender, partisanship) based on users' pre-invasion activities in other subreddits. We investigate relationships between users and their neighbors, finding a small correlation between user's sentiment and the average sentiment of their neighbors. Interestingly, we do not find evidence of homophilic interactions among specific demographic groups particularly engaged in discussions around the war.

\xhdr{Implications} Overall, our work provides new evidence about how the Russian invasion of Ukraine shaped the online debate in two European countries strongly involved in the support of the Ukrainian cause. We observed that users in these Reddit communities generally express a negative attitude towards the war. Notably, discussions about these events often exhibit a more toxic behavior. This suggests that a significant portion of users align their sentiments with the prevailing stance of their country regarding the Russian invasion of Ukraine. We also find that, in accordance with existing literature, Reddit interactions around the war did not lead to the formation of echo chambers.

\section{Related Work}
\label{sec:related_work}

There are a number of existing contributions that analyze the interplay between online conversations and the ongoing Russian invasion of Ukraine.

In March 2022, the Observatory on Social Media at Indiana University published a series of white papers \cite{OSMMarch2022,OSMMay2022} in which they investigate the presence of suspicious activity on Twitter during the first weeks of the invasion, showing distinct spikes in the creation of new accounts on Twitter.
Furthermore, the analysis reveals the existence of multiple coordinated clusters of accounts, encompassing a wide spectrum of behaviors including spam, promotional content, and even hate speech. The Social Media Lab at Toronto Metropolitan University deployed an online platform, called Russia-Ukraine ConflictMisinfo Research Portal \cite{SML_Toronto2022}, aimed to be a resource for researchers interested in studying the magnitude and the nature of online misinformation related to the conflict. They offer two main resources: the Russia-Ukraine ConflictMisinfo Dashboard, which is a tool where it's possible to track misinformation articles and fact-checking statements; the Russia-Ukraine ConflictMisinfo Geo-Map, a map where debunked claims about the conflict are shown with their specific geographical location.

\citet{A._2022} explored news coverage of the war in three different media ecosystems, Western, Russian and Chinese, by employing topic modeling and differential sentiment analysis. 
Their work finds important differences between the Western and the Chinese/Russian ecosystems, with the last two having a tighter reciprocal influence.
In another work \cite{Hanley_2022}, the same authors apply sentence-level topic analysis (MPNet model) to articles published by ten pro-Russian propaganda outlets, to understand the prominence of Russian state media narratives on selected subreddits.
They show that in r/Russia, almost 40\% of comments correspond to pro-Russian narratives, while in r/politics it is around 9\%.

Another line of research \cite{geissler2023russian,shen2023examining,boyte2017analysis,alyukov2023wartime} study the diffusion of Russian propaganda on social media and the role of social bots in the campaigns of information warfare.

\citet{pierri2023does} investigate the dynamics of account creation and suspension on Twitter during the Russian invasion of Ukraine and the 2022 French Presidential elections. The same authors 
carry out in \cite{pierri2023propaganda} a longitudinal study of the diffusion of misinformation about the Russian invasion of Ukraine which originated from the low-credibility sources posted on Facebook and Twitter and from Russian state outlets, during the first months of the conflict. 

Other research focuses on studying users' reactions to these events, such as the work by \citet{caprolu2023characterizing}, in which the authors analyze over 5 million tweets collected during the month before and after the beginning of the war highlighting abnormal patterns in users' sentiment. \citet{Guerra_2023} study how to measure fear and hope in war-related discussion on Reddit in 6 different European countries. They observed how hope spiked positively in case of important victories of the Ukrainian resistance and non-military
related events, while it spiked negatively in the correspondence of relevant strategic losses.
Lastly, we report several datasets published \cite{russo2023acti, Park_2022,haq2022twitter,fung2022weibo, Zhu_2022} over the last years which allow researchers and practitioners to study online conversations around the war on a variety of platforms such as Reddit, Twitter, Weibo, VKontakte, and Telegram.

\section{Methodology}
\label{sec:methodologies}
\subsection{Data Collection}

\begin{figure}[!t]
\centering
\includegraphics[width=\linewidth] {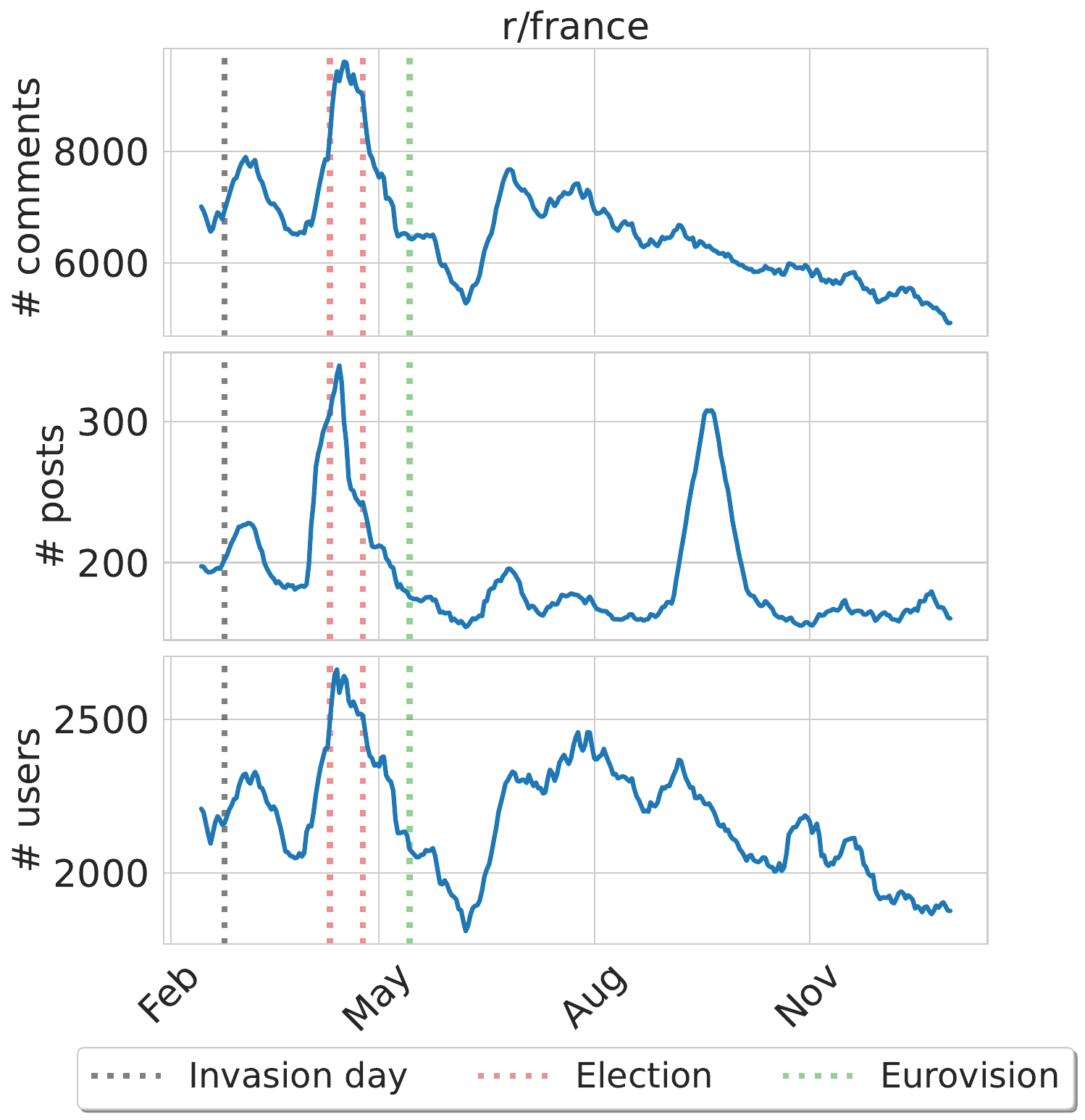}\label{fig:sent_1}

\caption{Time series showing the 7-day moving average of the daily number of comments, submissions and active users in \texttt{r/france}. }
\label{fig:france_descriptive}
\end{figure}

\begin{figure}[!t]
\centering

\includegraphics[width=\linewidth] {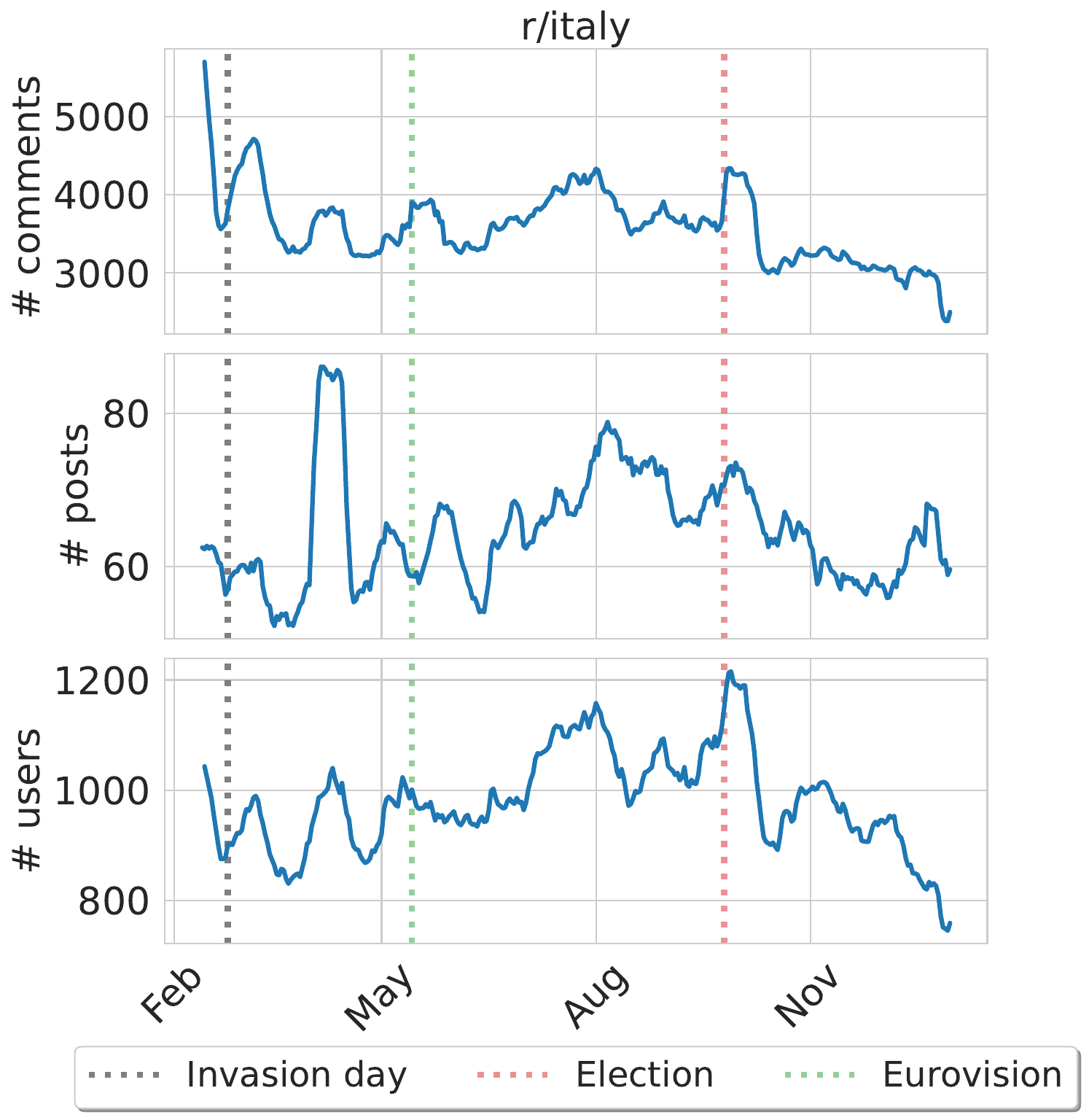}\label{fig:sent_2}
\caption{Time series showing the 7-day moving average of the daily number of comments, submissions and active users in \texttt{r/italy}. }
\label{fig:italy_descriptive}
\end{figure}

 To conduct our analysis, we collected all activity in the two subreddits during the period 01/02/2022 to 31/12/2022, leveraging the Pushshift dataset \cite{Baumgartner_2020}. The resulting data contains over 3M comments, in the local languages, and 80k submissions across the two subreddits. In Figures \ref{fig:france_descriptive} and \ref{fig:italy_descriptive} we show the temporal trends of activity of \texttt{r/france} and \texttt{r/italy} respectively. 
We see different volumes plotted in these figures, due to the difference in the user base of the communities (\texttt{r/france}: 600k, \texttt{r/italy} : 370k as of March 2022).
The number of daily comments for \texttt{r/france} ranges between 5k and 9k, while in \texttt{r/italy} it is between 3k and 5k. Similarly for daily submissions and active users, with \texttt{r/france} having an average of around 200 daily submissions and over 2k daily users, while \texttt{r/italy} shows an average of just over 60 daily submissions and 1k daily users.

In \texttt{r/france} we can observe the presence of a peak of daily comments, submissions and active users around the election period; this behavior is also found, with minor intensity, in \texttt{r/italy}.
This pattern is present in both communities around the beginning of the invasion, even though is smaller compared to the one for the elections.
To identify war-related comments, we applied a TF-IDF-based snowball sampling approach with keyword matching and manual inspection. This offers a valuable alternative to traditional probability sampling methods and it is widely employed in the literature \cite{russo2023spillover,deverna2021covaxxy,conover2012partisan}. We inspected a random sample,one for each subreddit, of 5.000 comments posted during the first month of the war that matched any of the following seed keywords. In Italian: ’Russia’, ’Ucraina’, ’Zelensky’, ’Putin’,’guerra’; in French: 'Russie','Ukraine','Putin','Zelensky','guerre'.
This way we could see what other meaningful words were used with known war-related keywords.

We then applied the previously defined sampling technique to obtain a final set of 20 relevant keywords, manually improved with case-sensitive alternatives to a total of 40, to be used to filter the body of all the comments in our dataset. 

We applied the same methodology both to the title and the body of the submissions, to identify those that were related to the Russian invasion of Ukraine and extract all the comments of these submissions.
This process yielded over 300k  war-related comments and 6.9k submissions between the two subreddits.

\subsection{Tracking moderators activity}

Reddit moderators are volunteer users who are tasked with the enactment of the subreddit rules and, more in general, to keep the community in order.So, moderators can remove comments that they deem to be against the rules of the community  \footnote{https://support.reddithelp.com/hc/en-us/articles/204533859-What-s-a-moderator - Accessed on 6/12/2023}.
In our dataset, removed comments are registered as a record with the body field equal to \texttt{[removed]} and the author
field equal to \texttt{[deleted]}, which implies that the content of removed comments cannot be inspected.
Since we cannot discern if a removed comment is war-related, we consider as war-related those that are submitted under a war-related post.
We identified a total of 26k removed comments in \texttt{r/italy} and 40k removed comments in \texttt{r/france}. 

\subsection{Sentiment and toxicity}
To assign a score of sentiment to Reddit comments, we employed the last available version, as of November 2023, of a multilingual BERT-based model by \citet{lik_xun_yuan_2023} for both Italian and French languages. To analyze toxicity, we relied on Detoxify \cite{Detoxify}, a BERT-based model that proved to be performing in both languages. 
We leveraged the \texttt{Transformers} library by Huggingface \cite{wolf-etal-2020-transformers} to deploy both models and to split the text of comments in our dataset into phrases we relied on the \texttt{NLTK} python library, obtaining a total of over 5.5M sentences.
For this analysis, we did not consider deleted and removed comments, since their textual content is not available.
We relied on the previously mentioned models and tools because of their state-of-the-art performance, popularity in the research community, and wide availability.

The sentiment analysis model outputs two numbers which indicate the probabilities for the text being negative and positive. To assign a continuous score, we consider the positive probability minus the negative probability, obtaining values in the range (-1, +1). We notice that we are not interested in the absolute value of the sentiment, but rather comparing the sentiment of online conversations across different periods and topics.
Toxicity scores are instead provided by Detoxify on a continuous scale between 0 and 1. If a comment is composed of multiple sentences, we take the average of sentiment and toxicity values.

As a reference for the sentiment and toxicity of online conversations, we consider three other important events that happened during the period of our analysis: Eurovision 2022 (10-14 May), the 2022 Italian political election (25 September), and the 2022 French political Election (10-24 April).

\subsection{Users' sentiment and socio-demographic scores}
Reddit is fundamentally different from social media platforms such as X (formerly Twitter), in which submissions naturally lead to interactions between users.
To build a network of interactions between Reddit users, we follow previous literature \cite{Cinelli_2021}: there is an interaction between User A and User B only if A has commented under a comment of User B. The resulting network is a directed graph where users are represented by nodes, and edges represent the action of commenting under another user's comment. We enrich our network by adding several node attributes: \texttt{Sentiment}, \texttt{Age}, \texttt{Gender}, and \texttt{Partisanship.}

Sentiment is computed as the average of sentiment scores of all comments posted by the user during the period of analysis. \texttt{Age}, \texttt{Gender} and Partisanship are based on the scores by \citet{Waller_2021}, and are obtained by analyzing the pre-invasion activity of users following an approach similar to \citet{Monti_2023}. Specifically, for user $i$ we define the social score $X_{i}$ as: 
\begin{equation}
\label{eq:scores}
    X_{i}=\frac{\sum_{j}^{n}v_{ij}X_{j}}{V_{i}}
\end{equation}
which corresponds to the weighted average of the score $X_{j}$ of all subreddits $j$ in which user $i$ has posted a comment during the period November 2021 to January 2022, weighted by the number of comments $v_{ij}$ posted by user $i$, normalized by the total number of comments $V_i$ posted by user $i$ during the same period.

Scores can be interpreted as the tendency of a user to visit subreddits that are usually populated by users with a given particular leaning. 
For example, r/teenagers has a younger population than r/RedditForGrownups. Another example is r/liberal versus r/conservative: the former will have a more left-leaning population than the latter. Scores are indicated as polarities:
For \texttt{Age}, a negative value means a young-leaning score while a positive indicates an old-leaning score.
For \texttt{Gender}, a negative value indicates a male-leaning score while a positive indicates a female-leaning score.
For political leaning, a negative value indicates more of a liberal-leaning score, while a positive value indicates a more conservative-leaning score.

\section{Results}
\label{sec:res_discs}

\subsection{Conversations around the war}
We first analyze the prevalence of Reddit comments related to the war. We observe similar patterns in the two communities: around the beginning of the invasion (grey dotted line) there is a peak of war-related comments, reaching over 20\% and 40\% of the daily comments posted in \texttt{r/france} and \texttt{r/italy}, respectively. These values reduce to an average of 10\% for \texttt{r/france} and around 5\% for \texttt{r/italy} for the following months.

\subsection{Patterns of content moderation}
\begin{figure}[!t]
\centering
\includegraphics[width=\linewidth]{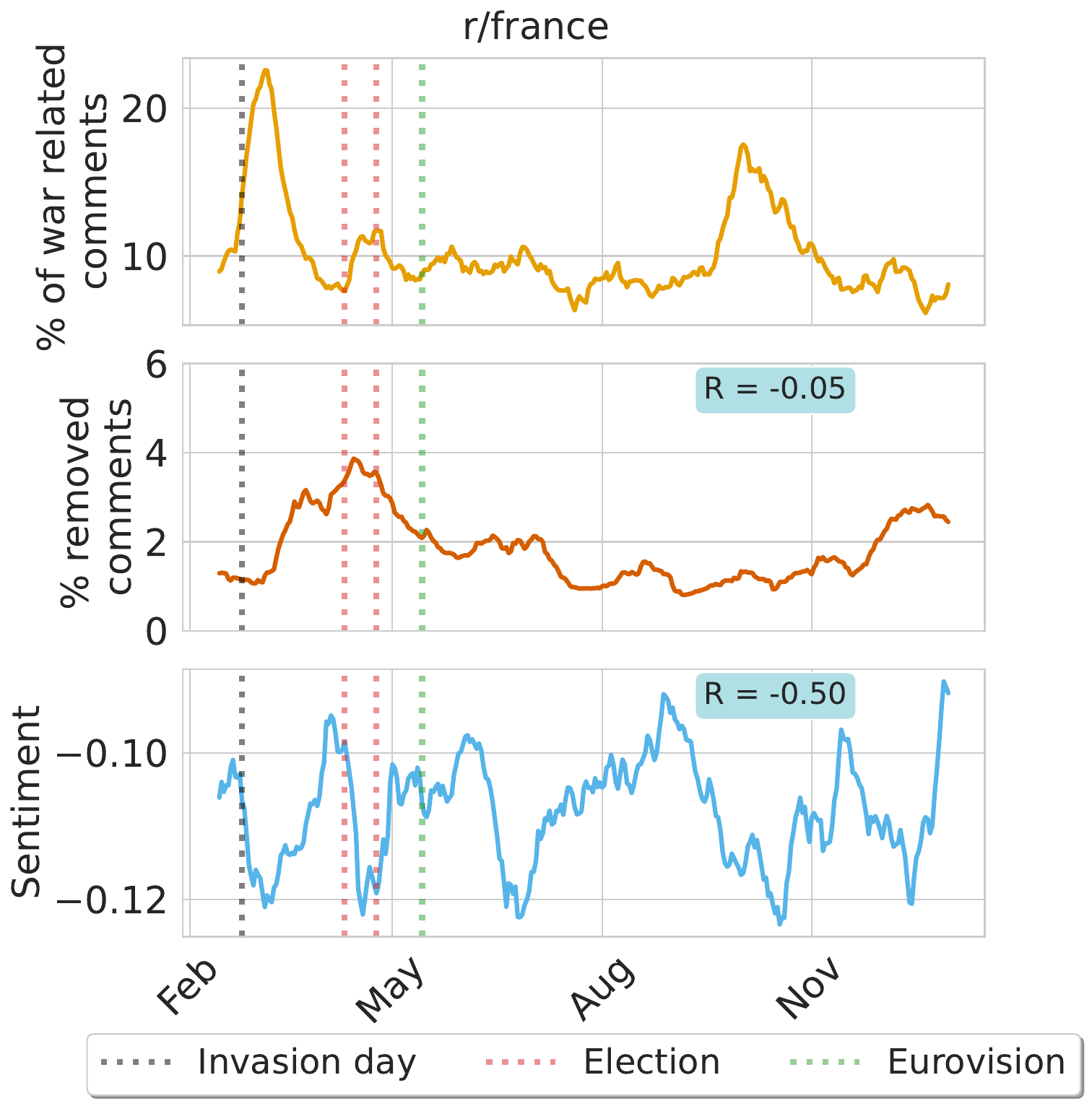}

\caption{Daily proportion of war-related comments (Top) and removed comments (Middle), and daily average sentiment (Bottom) for \texttt{r/france}. We show a 7-day moving average. We report the Pearson correlation (computed on the daily observations and not the moving average) between each time series and the proportion of war-related comments in the legend of each panel.}
\label{fig:all_times_fr}
\end{figure}

\begin{figure}[!t]
\centering
\includegraphics[width=\linewidth]{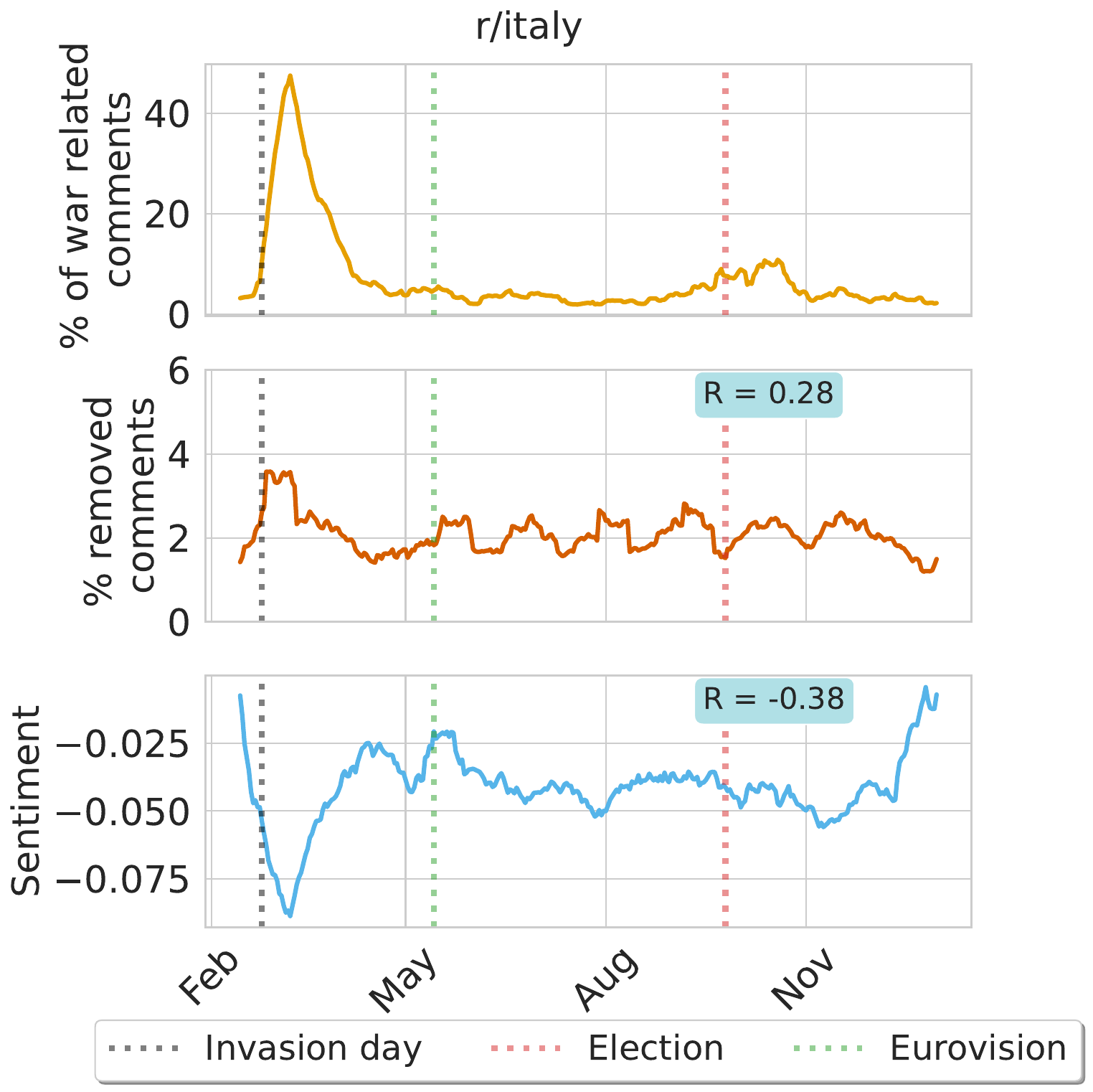}

\caption{Daily proportion of war-related comments (Top) and removed comments (Middle), and daily average sentiment (Bottom) for \texttt{r/italy}. We show a 7-day moving average. We report the Pearson correlation (computed on the daily observations and not the moving average) between each time series and the proportion of war-related comments in the legend of each panel.}
\label{fig:all_times_it}
\end{figure}

To answer our \textbf{(RQ1)}, we analyze
the activity of moderators in both subreddits during the period of analysis. We first observe a similar daily prevalence of removed comments in the two communities, as shown in the middle panel of Figures \ref{fig:all_times_fr} and \ref{fig:all_times_it}, with values in the range 2-4\%. However, we report different temporal patterns: In \texttt{r/france} we find a peak of removed comments in April, a month after the invasion, in correspondence with the 2022 Election. In \texttt{r/italy}, there is a peak of over 10\% daily removed comments (not shown in the Figure, which provides a 7-day moving average) in correspondence of a peak of war-related comments, i.e., the invasion day.
Consequently, we do not find a significant Pearson correlation between the prevalence of war-related comments and the daily percentage of removed comments in \texttt{r/france}, while we find a positive significant Pearson correlation (R=0.28, p<0.001) in  \texttt{r/Italy}.

\subsection{Sentiment and toxicity of war-related conversations}
To answer our \textbf{(RQ2)}, we first
compute the daily sentiment of conversations in the two subreddits by averaging the sentiment of all comments, for each day.
The bottom panels of Figures \ref{fig:all_times_fr} and \ref{fig:all_times_it} show that there is a significant negative correlation with the daily prevalence of war-related comments in both \texttt{r/france} and \texttt{r/italy}, with Pearson correlation coefficients respectively R=-0.50 and R=-0.38 (P < 0.001). In particular, we notice drops in the daily sentiment in correspondence with peaks of war-related comments. We also see the same behavior in \texttt{r/france} during the Election held in April. Lastly, we observe an increase in the daily sentiment towards the end of the year in both subreddits, likely due to conversations around Christmas and New Year's Eve.

To investigate whether different topics of conversation exhibit a different sentiment, we compare the daily values for four different cases, as described in the methods: war-related comments, election-related comments, comments about the Eurovision, and all other comments. We observe a general pattern which is common to the two subreddits: war-related comments are significantly more negative than all other comments. Besides, we observe that comments related to the Eurovision festival are much more positive than other conversations, and those related to the elections are more positive than war-related ones but less than Eurovision. All distributions are statistically different according to a two-way Mann-Whitney test (P < 0.001).

\begin{figure}[!t]
\centering
\includegraphics[width=\linewidth] {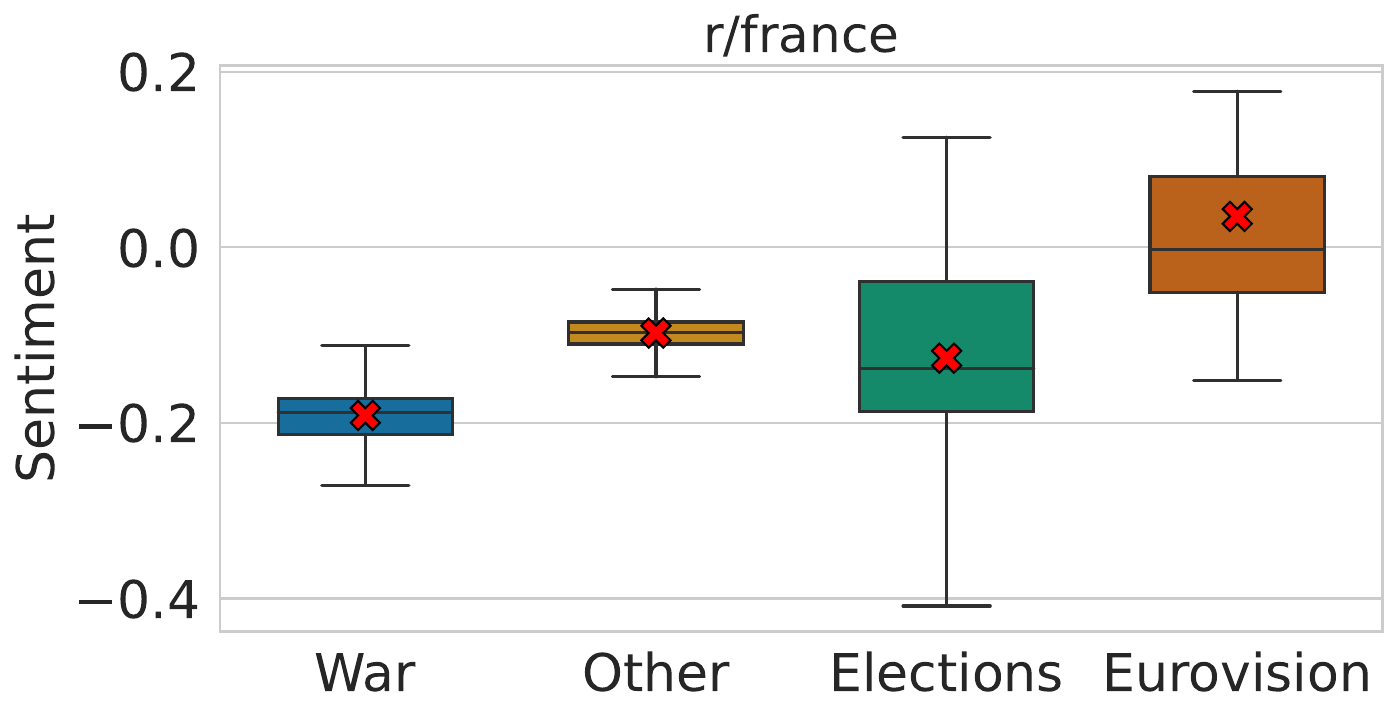}\label{fig:sent_1}
\includegraphics[width=\linewidth]{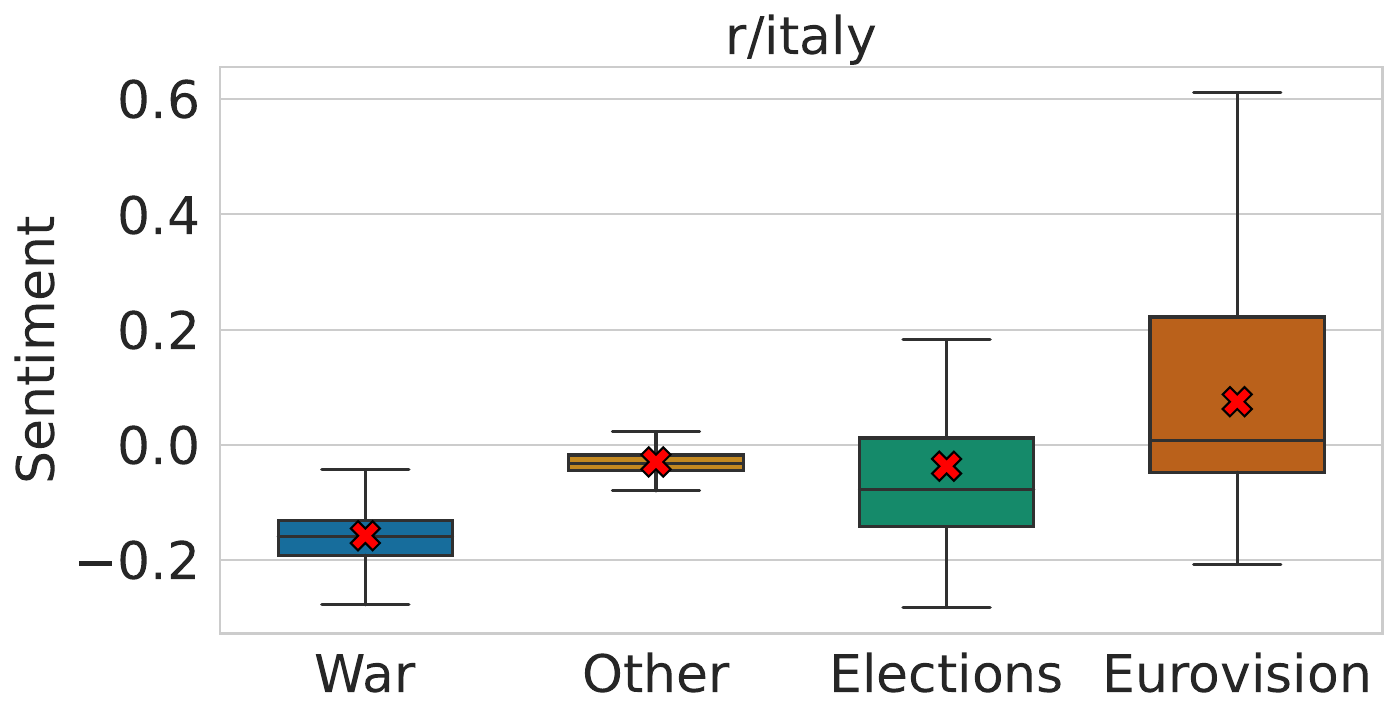}\label{fig:sent_2}
\caption{Distributions of the daily average sentiment in \texttt{r/france} (Top) and \texttt{r/italy} (Bottom) computed for non-overlapping subsets of comments corresponding to specific topics.
Outliers are not shown, and the red cross indicates the mean value of the distribution. Medians for \texttt{r/france} are:  War=-0.19, Other=-0.1, Elections=-0.14, Eurovision=-0.001. Medians for \texttt{r/italy} are: War=-0.16, Other=-0.03, Elections=-0.08, Eurovision=0.006.}
\label{fig:RQ2_daily_sentiment_distr}
\end{figure}

\begin{figure}[!t]
    \centering
    \includegraphics[width=\linewidth]{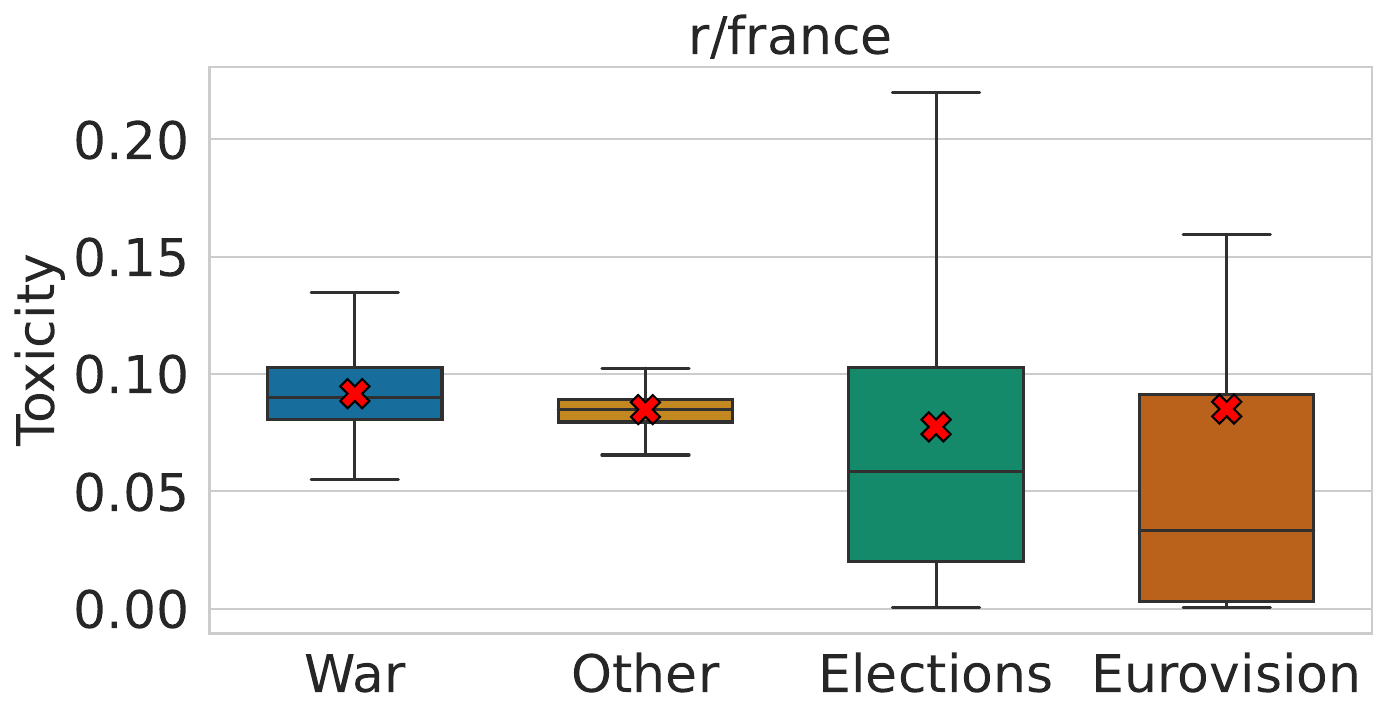}\label{fig:sub1}
    \includegraphics[width=\linewidth]{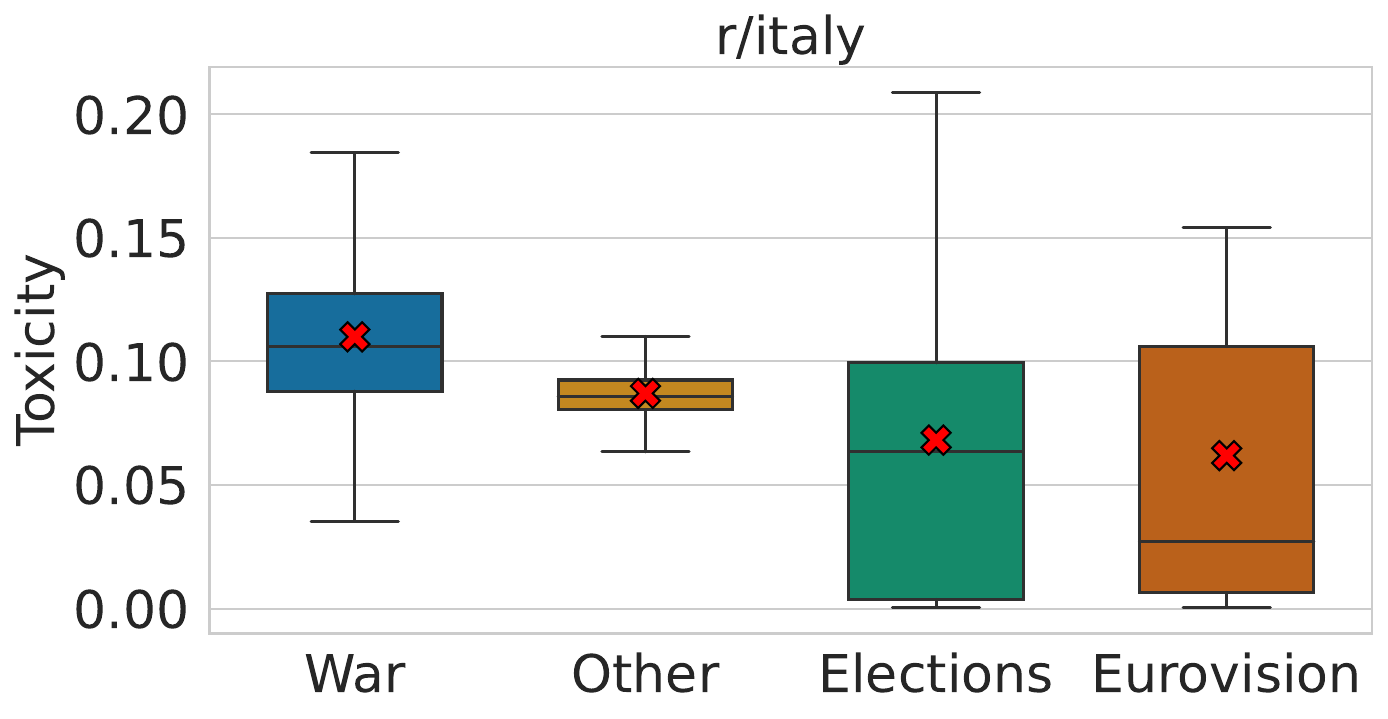}\label{fig:sub2}
    \caption{Distributions of the daily toxicity in \texttt{r/france} (Top) and \texttt{r/italy} (Bottom) computed for non-overlapping subsets of comments corresponding to specific topics. Outliers are not shown, and the red cross indicates the mean value of the distribution.Medians for \texttt{r/france} are:  War=0.091, Other=0.8, Elections=0.05, Eurovision=0.033. Medians for \texttt{r/italy} are: War=0.11, Other=0.085, 
    Elections=0.063, Eurovision=0.027.}
    \label{fig:RQ2_daily_tox_italy_distr}
\end{figure}

We replicate this analysis for the toxicity, finding no correlation and no interesting patterns for the daily average toxicity in the two subreddits (we omit the figure for the sake of brevity). Interestingly, we find discrepancies across topics indicating that toxicity is correlated with negative sentiment, as shown in Figure \ref{fig:RQ2_daily_tox_italy_distr}. Indeed, war-related comments are significantly more toxic than any other conversation in both subreddits. Comments related to Eurovision are the least toxic ones, with those about the elections being slightly less toxic than the average comment (cf. Other), although with a high variance in both subreddits which indicates the presence of outlier comments with an extreme toxicity. All distributions are significantly different according to a two-way Mann-Whitney test (P < 0.001), except for the case Eurovision versus Elections in \texttt{r/france} (P = 0.36).

\subsection{Homophily in war-related conversations}
\label{subsec:ans_rq3}
To answer our \textbf{(RQ3)}, we first analyze the network of interactions between users in the two subreddits, as described in the methodology.

\begin{figure}[!t]
\centering
{\includegraphics[width=\linewidth]{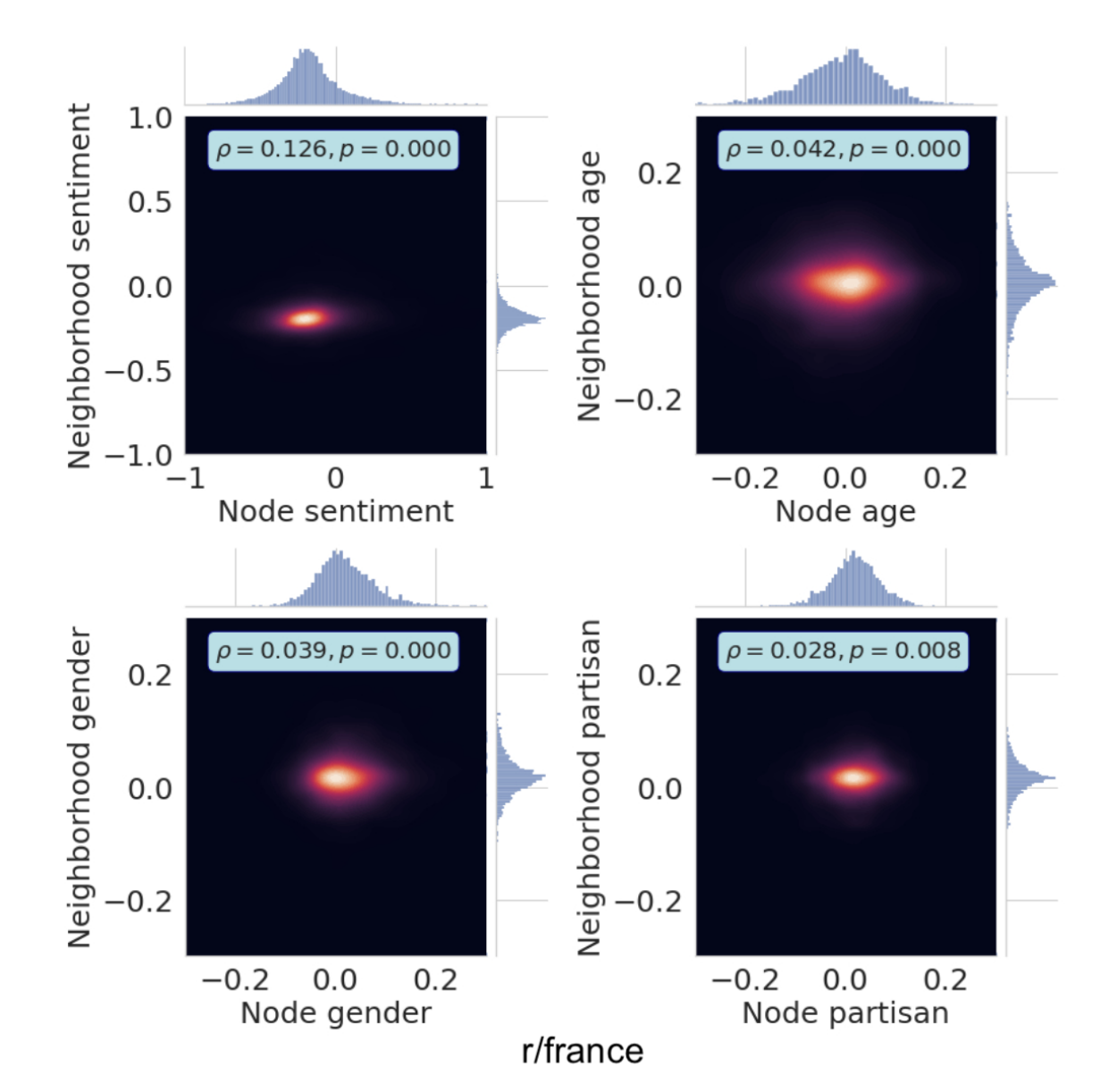}}

\caption{Joint distribution of the score of users (nodes) and the average score of their neighbors in \texttt{r/france}. \texttt{Sentiment} values show a weak significant correlation (Pearson R=0.126 P<0.001).}
\label{fig:RQ3_france_corr}

\end{figure}

\begin{figure}[!t]
\centering
{\includegraphics[width=\linewidth]{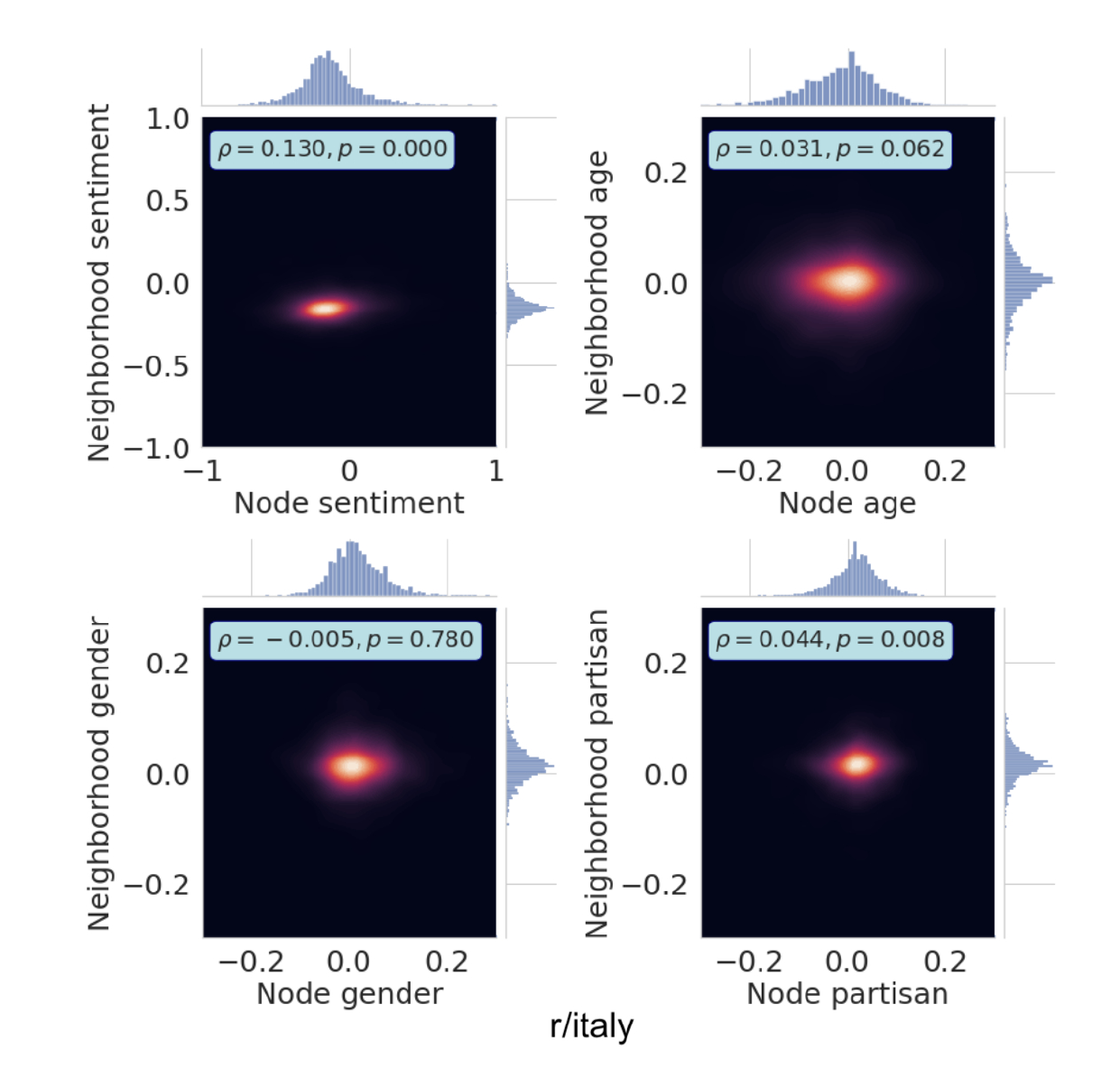}}
\caption{Joint distribution of the score of users (nodes) and the average score of their neighbors in \texttt{r/italy}. \texttt{Sentiment} values show a weak significant correlation (Pearson R=0.130,p<.001).}
\label{fig:RQ3_italy_corr}

\end{figure}

Following \citet{Cinelli_2021}, we test the presence of homophily between users according to the four user attributes: \texttt{Sentiment}, \texttt{Age}, \texttt{Gender}, \texttt{Partisanship}. Specifically, we study the joint distribution of users' attributes and the average value of their neighbors, as derived from the network of interactions. As shown in Figures \ref{fig:RQ3_france_corr} and \ref{fig:RQ3_italy_corr}, we do observe a weak significant correlation for what concerns the sentiment of users; this correlation is still significant for the socio-demographic scores but the coefficient is almost 0 (except for \texttt{Gender} in \texttt{r/Italy}, where P = 0.78). These results extend previous findings on the absence of political homophily and echo chambers in Reddit \cite{Cinelli_2021}, as we show the same pattern for \texttt{Age} and \texttt{Gender}.

\begin{figure}[!t]
\centering
\includegraphics[width=\linewidth]{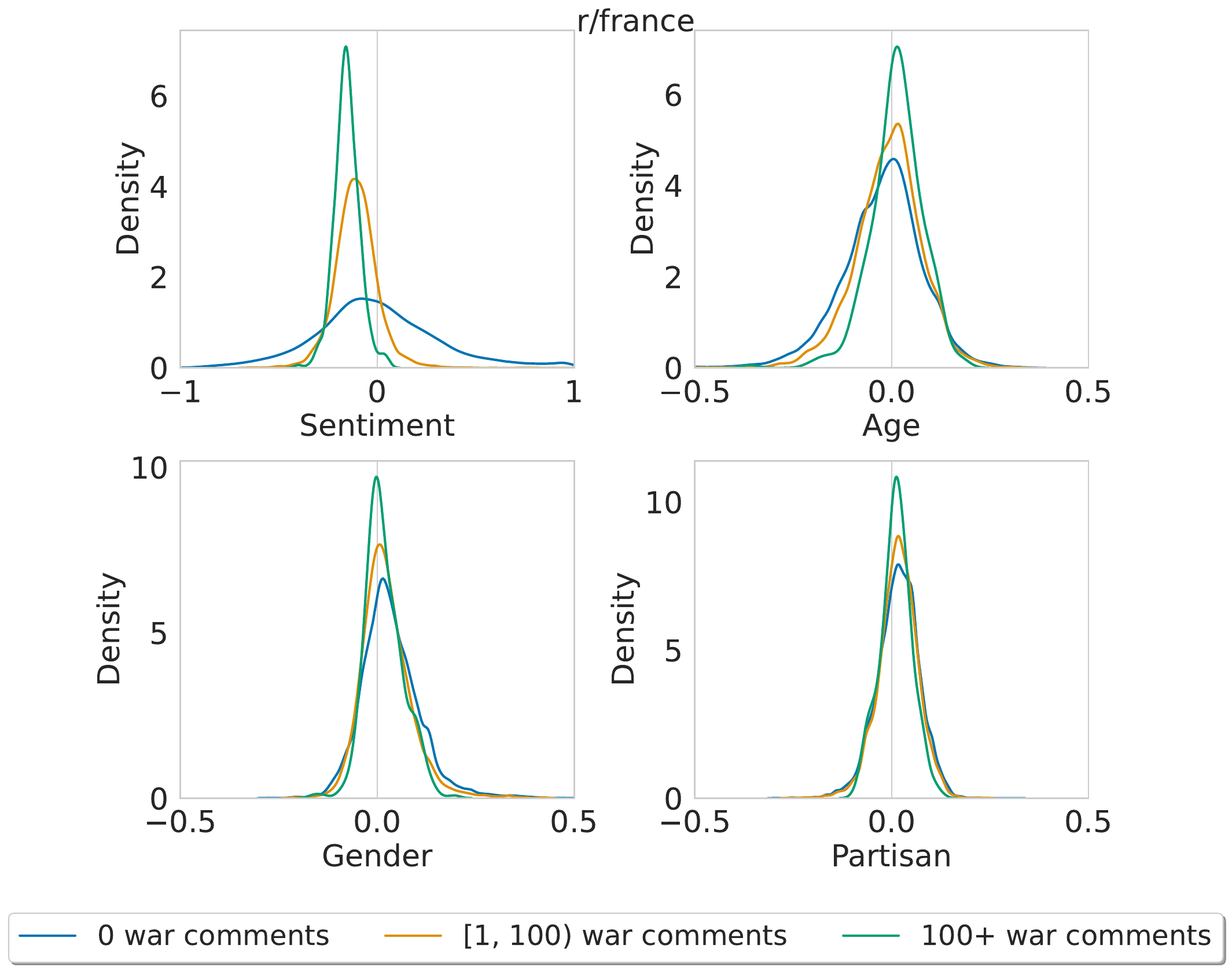}

\caption{Distributions of \texttt{Sentiment}, \texttt{Age}, \texttt{Gender} and Partisanship scores for users of \texttt{r/france} for the three classes of war-related activity.}
\label{fig:RQ3_all_part}
\end{figure}

\begin{figure}[!t]
\centering
\includegraphics[width=\linewidth]{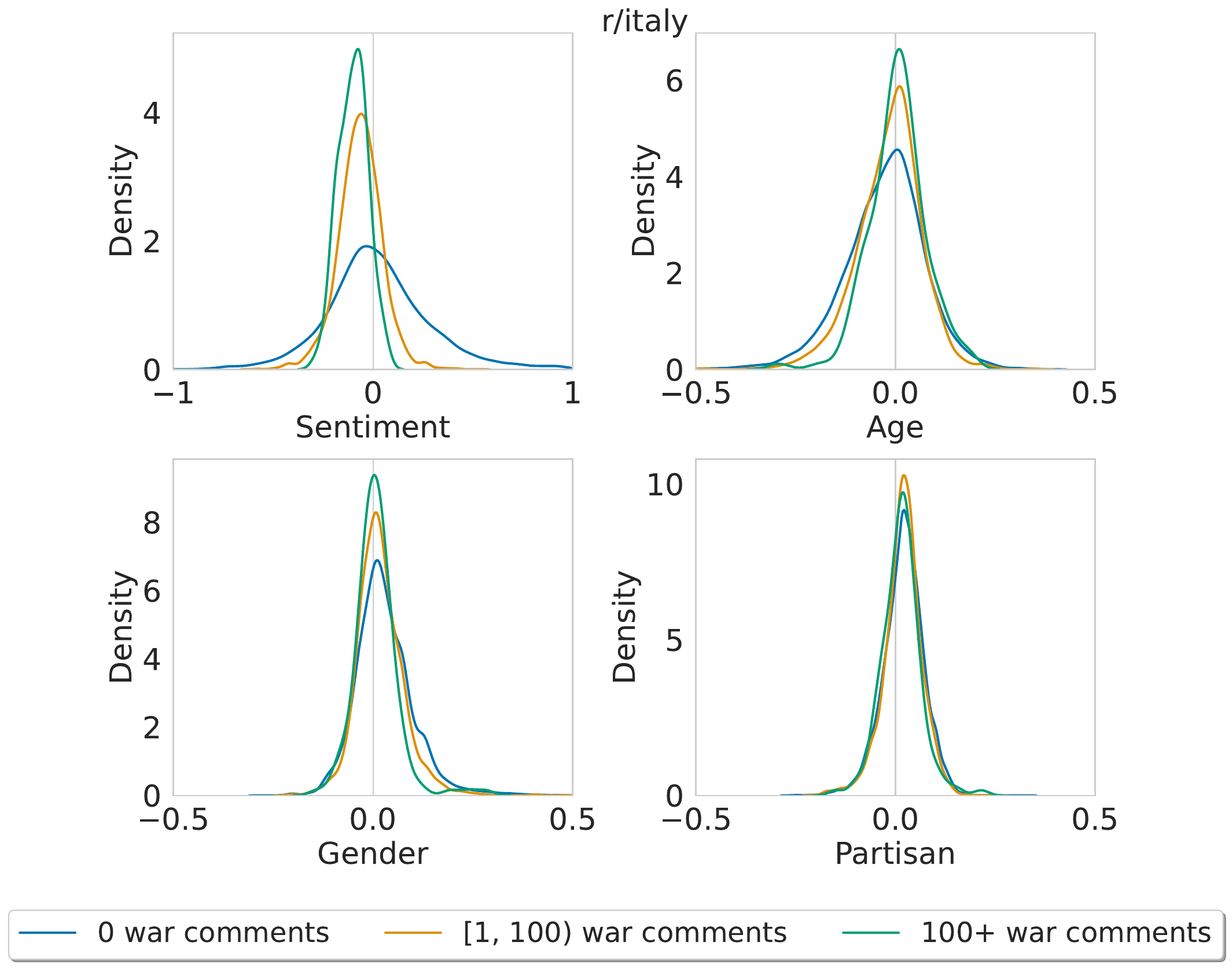}

\caption{Distributions of \texttt{Sentiment}, \texttt{Age}, \texttt{Gender} and \texttt{Partisanship} scores for users of \texttt{r/italy} for the three classes of war-related activity.}
\label{fig:RQ3_scores_italy}
\end{figure}

We further investigate whether users more active on the topic of war exhibit different attributes. In Figures \ref{fig:RQ3_all_part} and \ref{fig:RQ3_scores_italy} we show the distribution of all attributes for three different groups of users: those that never commented on the war, those with at least one and less than 100 war-related comments, and users with more than 100 war-related comments. We find a common pattern in the two subreddits: the \texttt{Age} median shifts from negative values to more positive ones, while the \texttt{Gender} median behaves in the opposite way, shifting from a more positive (female-leaning) to a more male-leaning one. Partisanship shifts from predominantly conservative-leaning scores to more central ones in both subreddits. Overall we can observe that users more engaged in war-related discussions show to be more male, older and center-oriented than the users who do not comment about the war. Despite being significant (Kruskal-Wallis test, P<0.05), however, the differences for the socio-demographic scores are very small (cf. Table \ref{tab:Medians}).

\begin{table}[]
\begin{tabular}{cccccc}
\hline
\multicolumn{1}{l}{}                           & \multicolumn{1}{l}{Activity} & \multicolumn{1}{l}{Sentiment} & \multicolumn{1}{l}{Age} & \multicolumn{1}{l}{Gender} & \multicolumn{1}{l}{Partisan} \\ \hline
\multicolumn{1}{c|}{\multirow{3}{*}{r/france}} & 0                            & -0.02                         & -0.016                  & 0.02                       & 0.019                        \\
\multicolumn{1}{c|}{}                          & {[}1,100)                    & -0.11                         & -0.002                  & 0.01                       & 0.017                        \\
\multicolumn{1}{c|}{}                          & 100+                         & -0.16                         & 0.015                   & 0.006                      & 0.012                        \\ \cline{2-6} 
\multicolumn{1}{c|}{\multirow{3}{*}{r/italy}}  & 0                            & 0.004                         & -0.02                   & 0.03                       & 0.021                        \\
\multicolumn{1}{c|}{}                          & {[}1,100)                    & -0.07                         & -0.007                  & 0.01                       & 0.019                        \\
\multicolumn{1}{c|}{}                          & 100+                         & -0.09                         & 0.006                   & 0.004                      & 0.013                       
\end{tabular}
\caption{Median values of the distribution of figures \ref{fig:RQ3_all_part},\ref{fig:RQ3_scores_italy} divided by subreddit and class of war related activity. All the medians between for each score in each subreddit are statistically different according to the Kruskal-Wallis test (p<0.05)}
\label{tab:Medians}
\end{table}

\section{Discussion}
\label{sec:conclusions}
\subsection{Contributions}
We conducted a longitudinal study, tracking eleven months of activity in the two most representative subreddits of Italian and French communities from the beginning of the Russian invasion of Ukraine to the end of 2022.
We identified over 200k comments and 10k submissions that discussed the invasion of Ukraine in both subreddits, highlighting how the war changed the moderation behavior in \texttt{r/italy}, while in \texttt{r/france} this effect was more visible during the Election.

We found a significant correlation between the prevalence of the war topic and the negativity of daily sentiment in comments shared by Reddit users in both communities; this difference was even more evident when compared to other significant events that happened during the time frame of our study (Eurovision 2022 and 2022 French and Italian Political Elections). We observed the same results when analyzing the toxicity of online conversations.

Finally, we found no evidence of political homophily on Reddit, as previously seen in the literature, extending this result to other socio-demographic scores and sentiment.

Our results prove that there are astonishing similarities between two representative Reddit communities of two of the most important European countries. Both highlighted a strong negativity towards the war, which is a shared sentiment with the alignment of the EU and respective countries. \footnotemark[2]

We also showed how the debate in these communities about this topic brought together a wide range of users with different demographic profiles.
We also showed some differences, for example in the moderation activity, shedding light on how the moderators of these communities perceive the importance of discussed topics in the community they administrate.

\subsection{Limitations}
There are a number of limitations to our study. 
First of all, even though we filtered out comments produced by known Reddit bots, such as u/RemindMeBot, u/savevideobot, we could not check for comments generated by inauthentic accounts.

Moreover, we used a keyword-matching-based method to identify comments related to the Russian invasion of Ukraine which tended to also signal which might entail false positives (i.e., generic comments about war). A manual validation on a random sample of 200 comments results in a small percentage (<10\%).

We analyzed sentiment and toxicity with black-box deep learning models that might not be completely accurate \cite{nogara2023toxic}.
We sampled 200 random sentences (100 for each subreddit) and manually validated the classifiers, finding a small percentage of inaccurate results (<10\%). 

Finally, only 56\% of the users in our study were also active in the three months before February 2022 ( necessary to obtain social scores) with the remaining 44\% joining the subreddits in the following months.
We did not notice any abnormal pattern around the way these new users joined the communities.

Furthermore, the scores we used for describing the political leaning of the users are referring more to a US-oriented political system than the Italian or French system.
\subsection{Future Work}
There are several opportunities for future work related to our research and this topic in general.
It would be relevant to conduct the same analysis on other European subreddits, maybe of the countries geographically closer to Ukraine,  to highlight eventual differences or points of similarity with the ones we analyzed in this study.
Our results could also be compared with X (formerly Twitter) or Facebook, which are fundamentally different platforms and could highlight structural differences that could improve or worsen the discussion on these controversial topics.

\subsection{Ethical Considerations}
 In this study, we gave utmost attention to the principles of privacy, transparency, and fairness. Data privacy concerns are addressed through rigorous anonymization and aggregation procedures, without any attempt of backtracking to the real user. We employed transparent methodologies, detailing data collection, preprocessing steps, and modeling choices, enabling reproducibility. The study is committed to ensuring fairness in the analysis and  of results. We acknowledge the unavailability of the Reddit API, so we publish in an online repository the IDs of the comments that we employed in our dataset along with the Python notebooks we used to produce the results. The dataset can be reconstructed following the instructions provided in the description of the repository. \footnote{https://github.com/orsoFra/LS\_FRIT\_UKR}

\bibliographystyle{ACM-Reference-Format}
\bibliography{main}


\begin{thebibliography}{32}


\ifx \showCODEN    \undefined \def \showCODEN     #1{\unskip}     \fi
\ifx \showDOI      \undefined \def \showDOI       #1{#1}\fi
\ifx \showISBNx    \undefined \def \showISBNx     #1{\unskip}     \fi
\ifx \showISBNxiii \undefined \def \showISBNxiii  #1{\unskip}     \fi
\ifx \showISSN     \undefined \def \showISSN      #1{\unskip}     \fi
\ifx \showLCCN     \undefined \def \showLCCN      #1{\unskip}     \fi
\ifx \shownote     \undefined \def \shownote      #1{#1}          \fi
\ifx \showarticletitle \undefined \def \showarticletitle #1{#1}   \fi
\ifx \showURL      \undefined \def \showURL       {\relax}        \fi
\providecommand\bibfield[2]{#2}
\providecommand\bibinfo[2]{#2}
\providecommand\natexlab[1]{#1}
\providecommand\showeprint[2][]{arXiv:#2}

\bibitem[Alyukov et~al\mbox{.}(2023)]%
        {alyukov2023wartime}
\bibfield{author}{\bibinfo{person}{Maxim Alyukov}, \bibinfo{person}{Maria Kunilovskaya}, {and} \bibinfo{person}{Andrei Semenov}.} \bibinfo{year}{2023}\natexlab{}.
\newblock \showarticletitle{Wartime Media Monitor (WarMM-2022): A Study of Information Manipulation on Russian Social Media during the Russia-Ukraine War}. In \bibinfo{booktitle}{\emph{Proceedings of the 7th Joint SIGHUM Workshop on Computational Linguistics for Cultural Heritage, Social Sciences, Humanities and Literature}}. \bibinfo{pages}{152--161}.
\newblock


\bibitem[Baumgartner et~al\mbox{.}(2020)]%
        {Baumgartner_2020}
\bibfield{author}{\bibinfo{person}{Jason Baumgartner}, \bibinfo{person}{Savvas Zannettou}, \bibinfo{person}{Brian Keegan}, \bibinfo{person}{Megan Squire}, {and} \bibinfo{person}{Jeremy Blackburn}.} \bibinfo{year}{2020}\natexlab{}.
\newblock \showarticletitle{The Pushshift Reddit Dataset}.
\newblock \bibinfo{journal}{\emph{International Conference on Web and Social Media}} (\bibinfo{year}{2020}).
\newblock
\urldef\tempurl%
\url{https://doi.org/10.5281/zenodo.3608135}
\showDOI{\tempurl}


\bibitem[Boyte(2017)]%
        {boyte2017analysis}
\bibfield{author}{\bibinfo{person}{Kenneth~J Boyte}.} \bibinfo{year}{2017}\natexlab{}.
\newblock \showarticletitle{An analysis of the social-media technology, tactics, and narratives used to control perception in the propaganda war over Ukraine}.
\newblock \bibinfo{journal}{\emph{Journal of Information Warfare}} \bibinfo{volume}{16}, \bibinfo{number}{1} (\bibinfo{year}{2017}), \bibinfo{pages}{88--111}.
\newblock


\bibitem[Caprolu et~al\mbox{.}(2023)]%
        {caprolu2023characterizing}
\bibfield{author}{\bibinfo{person}{Maurantonio Caprolu}, \bibinfo{person}{Alireza Sadighian}, {and} \bibinfo{person}{Roberto Di~Pietro}.} \bibinfo{year}{2023}\natexlab{}.
\newblock \showarticletitle{Characterizing the 2022-Russo-Ukrainian Conflict Through the Lenses of Aspect-Based Sentiment Analysis: Dataset, Methodology, and Key Findings}. In \bibinfo{booktitle}{\emph{2023 32nd International Conference on Computer Communications and Networks (ICCCN)}}. IEEE, \bibinfo{pages}{1--10}.
\newblock


\bibitem[Cinelli et~al\mbox{.}(2021)]%
        {Cinelli_2021}
\bibfield{author}{\bibinfo{person}{Matteo Cinelli}, \bibinfo{person}{Gianmarco De~Francisci Morales}, \bibinfo{person}{Alessandro Galeazzi}, \bibinfo{person}{Walter Quattrociocchi}, {and} \bibinfo{person}{Michele Starnini}.} \bibinfo{year}{2021}\natexlab{}.
\newblock \showarticletitle{The echo chamber effect on social media}.
\newblock \bibinfo{journal}{\emph{Proceedings of the National Academy of Sciences of the United States of America}} (\bibinfo{year}{2021}).
\newblock
\urldef\tempurl%
\url{https://doi.org/10.1073/pnas.2023301118}
\showDOI{\tempurl}


\bibitem[Conover et~al\mbox{.}(2012)]%
        {conover2012partisan}
\bibfield{author}{\bibinfo{person}{Michael~D Conover}, \bibinfo{person}{Bruno Gon{\c{c}}alves}, \bibinfo{person}{Alessandro Flammini}, {and} \bibinfo{person}{Filippo Menczer}.} \bibinfo{year}{2012}\natexlab{}.
\newblock \showarticletitle{Partisan asymmetries in online political activity}.
\newblock \bibinfo{journal}{\emph{EPJ Data science}} \bibinfo{volume}{1}, \bibinfo{number}{1} (\bibinfo{year}{2012}), \bibinfo{pages}{1--19}.
\newblock


\bibitem[DeVerna et~al\mbox{.}(2021)]%
        {deverna2021covaxxy}
\bibfield{author}{\bibinfo{person}{Matthew~R DeVerna}, \bibinfo{person}{Francesco Pierri}, \bibinfo{person}{Bao~Tran Truong}, \bibinfo{person}{John Bollenbacher}, \bibinfo{person}{David Axelrod}, \bibinfo{person}{Niklas Loynes}, \bibinfo{person}{Christopher Torres-Lugo}, \bibinfo{person}{Kai-Cheng Yang}, \bibinfo{person}{Filippo Menczer}, {and} \bibinfo{person}{John Bryden}.} \bibinfo{year}{2021}\natexlab{}.
\newblock \showarticletitle{CoVaxxy: A collection of English-language Twitter posts about COVID-19 vaccines}. In \bibinfo{booktitle}{\emph{Proceedings of the International AAAI Conference on Web and Social Media}}, Vol.~\bibinfo{volume}{15}. \bibinfo{pages}{992--999}.
\newblock


\bibitem[Di~Giovanni et~al\mbox{.}(2022)]%
        {di2022vaccineu}
\bibfield{author}{\bibinfo{person}{Marco Di~Giovanni}, \bibinfo{person}{Francesco Pierri}, \bibinfo{person}{Christopher Torres-Lugo}, {and} \bibinfo{person}{Marco Brambilla}.} \bibinfo{year}{2022}\natexlab{}.
\newblock \showarticletitle{VaccinEU: COVID-19 vaccine conversations on Twitter in French, German and Italian}. In \bibinfo{booktitle}{\emph{Proceedings of the International AAAI Conference on Web and Social Media}}, Vol.~\bibinfo{volume}{16}. \bibinfo{pages}{1236--1244}.
\newblock


\bibitem[Fung and Ji(2022)]%
        {fung2022weibo}
\bibfield{author}{\bibinfo{person}{Yi~R Fung} {and} \bibinfo{person}{Heng Ji}.} \bibinfo{year}{2022}\natexlab{}.
\newblock \showarticletitle{A weibo dataset for the 2022 russo-ukrainian crisis}.
\newblock \bibinfo{journal}{\emph{arXiv preprint arXiv:2203.05967}} (\bibinfo{year}{2022}).
\newblock


\bibitem[Geissler et~al\mbox{.}(2023)]%
        {geissler2023russian}
\bibfield{author}{\bibinfo{person}{Dominique Geissler}, \bibinfo{person}{Dominik B{\"a}r}, \bibinfo{person}{Nicolas Pr{\"o}llochs}, {and} \bibinfo{person}{Stefan Feuerriegel}.} \bibinfo{year}{2023}\natexlab{}.
\newblock \showarticletitle{Russian propaganda on social media during the 2022 invasion of Ukraine}.
\newblock \bibinfo{journal}{\emph{EPJ Data Science}} \bibinfo{volume}{12}, \bibinfo{number}{1} (\bibinfo{year}{2023}), \bibinfo{pages}{35}.
\newblock


\bibitem[Guerra and Karakus(2023)]%
        {Guerra_2023}
\bibfield{author}{\bibinfo{person}{Alessio Guerra} {and} \bibinfo{person}{Oktay Karakus}.} \bibinfo{year}{2023}\natexlab{}.
\newblock \showarticletitle{Sentiment analysis for measuring hope and fear from Reddit posts during the 2022 Russo-Ukrainian conflict}.
\newblock \bibinfo{journal}{\emph{Frontiers in Artificial Intelligence}} (\bibinfo{year}{2023}).
\newblock
\urldef\tempurl%
\url{https://doi.org/10.3389/frai.2023.1163577}
\showDOI{\tempurl}


\bibitem[Hanley et~al\mbox{.}(2023a)]%
        {A._2022}
\bibfield{author}{\bibinfo{person}{Hans~WA Hanley}, \bibinfo{person}{Deepak Kumar}, {and} \bibinfo{person}{Zakir Durumeric}.} \bibinfo{year}{2023}\natexlab{a}.
\newblock \showarticletitle{" A Special Operation": A Quantitative Approach to Dissecting and Comparing Different Media Ecosystems’ Coverage of the Russo-Ukrainian War}. In \bibinfo{booktitle}{\emph{Proceedings of the International AAAI Conference on Web and Social Media}}, Vol.~\bibinfo{volume}{17}. \bibinfo{pages}{339--350}.
\newblock


\bibitem[Hanley et~al\mbox{.}(2023b)]%
        {Hanley_2022}
\bibfield{author}{\bibinfo{person}{Hans~WA Hanley}, \bibinfo{person}{Deepak Kumar}, {and} \bibinfo{person}{Zakir Durumeric}.} \bibinfo{year}{2023}\natexlab{b}.
\newblock \showarticletitle{Happenstance: Utilizing Semantic Search to Track Russian State Media Narratives about the Russo-Ukrainian War On Reddit}. In \bibinfo{booktitle}{\emph{Proceedings of the international AAAI conference on web and social media}}, Vol.~\bibinfo{volume}{17}. \bibinfo{pages}{327--338}.
\newblock


\bibitem[Hanu and {Unitary team}(2020)]%
        {Detoxify}
\bibfield{author}{\bibinfo{person}{Laura Hanu} {and} \bibinfo{person}{{Unitary team}}.} \bibinfo{year}{2020}\natexlab{}.
\newblock \bibinfo{title}{Detoxify}.
\newblock \bibinfo{howpublished}{Github. https://github.com/unitaryai/detoxify}.
\newblock


\bibitem[Haq et~al\mbox{.}(2022)]%
        {haq2022twitter}
\bibfield{author}{\bibinfo{person}{Ehsan-Ul Haq}, \bibinfo{person}{Gareth Tyson}, \bibinfo{person}{Lik-Hang Lee}, \bibinfo{person}{Tristan Braud}, {and} \bibinfo{person}{Pan Hui}.} \bibinfo{year}{2022}\natexlab{}.
\newblock \showarticletitle{Twitter dataset for 2022 russo-ukrainian crisis}.
\newblock \bibinfo{journal}{\emph{arXiv preprint arXiv:2203.02955}} (\bibinfo{year}{2022}).
\newblock


\bibitem[{Lik Xun Yuan}(2023)]%
        {lik_xun_yuan_2023}
\bibfield{author}{\bibinfo{person}{{Lik Xun Yuan}}.} \bibinfo{year}{2023}\natexlab{}.
\newblock \bibinfo{title}{distilbert-base-multilingual-cased-sentiments-student (Revision 2e33845)}.
\newblock
\newblock
\urldef\tempurl%
\url{https://doi.org/10.57967/hf/1422}
\showDOI{\tempurl}


\bibitem[Monti et~al\mbox{.}(2023)]%
        {Monti_2023}
\bibfield{author}{\bibinfo{person}{Corrado Monti}, \bibinfo{person}{Jacopo D'Ignazi}, \bibinfo{person}{Michele Starnini}, {and} \bibinfo{person}{Gianmarco De~Francisci Morales}.} \bibinfo{year}{2023}\natexlab{}.
\newblock \showarticletitle{Evidence of Demographic rather than Ideological Segregation in News Discussion on Reddit}.
\newblock \bibinfo{journal}{\emph{The Web Conference}} (\bibinfo{year}{2023}).
\newblock
\urldef\tempurl%
\url{https://doi.org/10.1145/3543507.3583468}
\showDOI{\tempurl}


\bibitem[Nogara et~al\mbox{.}(2023)]%
        {nogara2023toxic}
\bibfield{author}{\bibinfo{person}{Gianluca Nogara}, \bibinfo{person}{Francesco Pierri}, \bibinfo{person}{Stefano Cresci}, \bibinfo{person}{Luca Luceri}, \bibinfo{person}{Petter T{\"o}rnberg}, {and} \bibinfo{person}{Silvia Giordano}.} \bibinfo{year}{2023}\natexlab{}.
\newblock \showarticletitle{Toxic Bias: Perspective API misreads German as more toxic}.
\newblock \bibinfo{journal}{\emph{arXiv preprint arXiv:2312.12651}} (\bibinfo{year}{2023}).
\newblock


\bibitem[OSM(2022a)]%
        {OSMMay2022}
\bibfield{author}{\bibinfo{person}{OSM}.} \bibinfo{year}{2022}\natexlab{a}.
\newblock \bibinfo{booktitle}{\emph{Analysis of Twitter accounts created around the invasion of Ukraine}}.
\newblock
\urldef\tempurl%
\url{https://osome.iu.edu/research/white-papers/Ukraine_OSoMe_White_Paper_May_2022.pdf}
\showURL{%
\tempurl}


\bibitem[OSM(2022b)]%
        {OSMMarch2022}
\bibfield{author}{\bibinfo{person}{OSM}.} \bibinfo{year}{2022}\natexlab{b}.
\newblock \bibinfo{booktitle}{\emph{Suspicious Twitter Activity around the Russian Invasion of Ukraine}}.
\newblock
\urldef\tempurl%
\url{https://osome.iu.edu/research/white-papers/Ukraine_OSoMe_White_Paper_March_2022.pdf}
\showURL{%
\tempurl}


\bibitem[Park et~al\mbox{.}(2022)]%
        {Park_2022}
\bibfield{author}{\bibinfo{person}{Chan~Young Park}, \bibinfo{person}{Julia Mendelsohn}, \bibinfo{person}{Anjalie Field}, {and} \bibinfo{person}{Yulia Tsvetkov}.} \bibinfo{year}{2022}\natexlab{}.
\newblock \showarticletitle{Challenges and Opportunities in Information Manipulation Detection: An Examination of Wartime Russian Media}.
\newblock \bibinfo{journal}{\emph{Conference on Empirical Methods in Natural Language Processing}} (\bibinfo{year}{2022}).
\newblock
\urldef\tempurl%
\url{https://doi.org/null}
\showDOI{\tempurl}


\bibitem[Pierri(2020)]%
        {pierri2020diffusion}
\bibfield{author}{\bibinfo{person}{Francesco Pierri}.} \bibinfo{year}{2020}\natexlab{}.
\newblock \showarticletitle{The diffusion of mainstream and disinformation news on Twitter: the case of Italy and France}. In \bibinfo{booktitle}{\emph{Companion proceedings of the web conference 2020}}. \bibinfo{pages}{617--622}.
\newblock


\bibitem[Pierri et~al\mbox{.}(2023a)]%
        {pierri2023does}
\bibfield{author}{\bibinfo{person}{Francesco Pierri}, \bibinfo{person}{Luca Luceri}, \bibinfo{person}{Emily Chen}, {and} \bibinfo{person}{Emilio Ferrara}.} \bibinfo{year}{2023}\natexlab{a}.
\newblock \showarticletitle{How does Twitter account moderation work? Dynamics of account creation and suspension on Twitter during major geopolitical events}.
\newblock \bibinfo{journal}{\emph{EPJ Data Science}} \bibinfo{volume}{12}, \bibinfo{number}{1} (\bibinfo{year}{2023}), \bibinfo{pages}{43}.
\newblock


\bibitem[Pierri et~al\mbox{.}(2023b)]%
        {pierri2023propaganda}
\bibfield{author}{\bibinfo{person}{Francesco Pierri}, \bibinfo{person}{Luca Luceri}, \bibinfo{person}{Nikhil Jindal}, {and} \bibinfo{person}{Emilio Ferrara}.} \bibinfo{year}{2023}\natexlab{b}.
\newblock \showarticletitle{Propaganda and Misinformation on Facebook and Twitter during the Russian Invasion of Ukraine}. In \bibinfo{booktitle}{\emph{Proceedings of the 15th ACM Web Science Conference 2023}}. \bibinfo{pages}{65--74}.
\newblock


\bibitem[Proedrou(2010)]%
        {proedrou2010ukraine}
\bibfield{author}{\bibinfo{person}{Filippos Proedrou}.} \bibinfo{year}{2010}\natexlab{}.
\newblock \showarticletitle{Ukraine’s foreign policy: accounting for Ukraine’s indeterminate stance between Russia and the West}.
\newblock \bibinfo{journal}{\emph{Southeast European and Black Sea Studies}} \bibinfo{volume}{10}, \bibinfo{number}{4} (\bibinfo{year}{2010}), \bibinfo{pages}{443--456}.
\newblock


\bibitem[Russo et~al\mbox{.}(2023a)]%
        {russo2023acti}
\bibfield{author}{\bibinfo{person}{Giuseppe Russo}, \bibinfo{person}{Niklas Stoehr}, {and} \bibinfo{person}{Manoel~Horta Ribeiro}.} \bibinfo{year}{2023}\natexlab{a}.
\newblock \showarticletitle{Acti at evalita 2023: Overview of the conspiracy theory identification task}.
\newblock \bibinfo{journal}{\emph{arXiv preprint arXiv:2307.06954}} (\bibinfo{year}{2023}).
\newblock


\bibitem[Russo et~al\mbox{.}(2023b)]%
        {russo2023spillover}
\bibfield{author}{\bibinfo{person}{Giuseppe Russo}, \bibinfo{person}{Luca Verginer}, \bibinfo{person}{Manoel~Horta Ribeiro}, {and} \bibinfo{person}{Giona Casiraghi}.} \bibinfo{year}{2023}\natexlab{b}.
\newblock \showarticletitle{Spillover of antisocial behavior from fringe platforms: The unintended consequences of community banning}. In \bibinfo{booktitle}{\emph{Proceedings of the International AAAI Conference on Web and Social Media}}, Vol.~\bibinfo{volume}{17}. \bibinfo{pages}{742--753}.
\newblock


\bibitem[Shen et~al\mbox{.}(2023)]%
        {shen2023examining}
\bibfield{author}{\bibinfo{person}{Fei Shen}, \bibinfo{person}{Erkun Zhang}, \bibinfo{person}{Hongzhong Zhang}, \bibinfo{person}{Wujiong Ren}, \bibinfo{person}{Quanxin Jia}, {and} \bibinfo{person}{Yuan He}.} \bibinfo{year}{2023}\natexlab{}.
\newblock \showarticletitle{Examining the differences between human and bot social media accounts: A case study of the Russia-Ukraine War}.
\newblock \bibinfo{journal}{\emph{First Monday}} (\bibinfo{year}{2023}).
\newblock


\bibitem[University(2022)]%
        {SML_Toronto2022}
\bibfield{author}{\bibinfo{person}{Social Media Lab Toronto~Metropolitan University}.} \bibinfo{year}{2022}\natexlab{}.
\newblock \bibinfo{booktitle}{\emph{Russia-Ukraine ConflictMisinfo Research Portal}}.
\newblock
\urldef\tempurl%
\url{https://conflictmisinfo.org}
\showURL{%
\tempurl}


\bibitem[Waller and Anderson(2021)]%
        {Waller_2021}
\bibfield{author}{\bibinfo{person}{Isaac Waller} {and} \bibinfo{person}{Ashton Anderson}.} \bibinfo{year}{2021}\natexlab{}.
\newblock \showarticletitle{Quantifying social organization and political polarization in online platforms}.
\newblock \bibinfo{journal}{\emph{Nature}} (\bibinfo{year}{2021}).
\newblock
\urldef\tempurl%
\url{https://doi.org/10.1038/s41586-021-04167-x}
\showDOI{\tempurl}


\bibitem[Wolf et~al\mbox{.}(2020)]%
        {wolf-etal-2020-transformers}
\bibfield{author}{\bibinfo{person}{Thomas Wolf}, \bibinfo{person}{Lysandre Debut}, \bibinfo{person}{Victor Sanh}, \bibinfo{person}{Julien Chaumond}, \bibinfo{person}{Clement Delangue}, \bibinfo{person}{Anthony Moi}, \bibinfo{person}{Pierric Cistac}, \bibinfo{person}{Tim Rault}, \bibinfo{person}{Remi Louf}, \bibinfo{person}{Morgan Funtowicz}, \bibinfo{person}{Joe Davison}, \bibinfo{person}{Sam Shleifer}, \bibinfo{person}{Patrick von Platen}, \bibinfo{person}{Clara Ma}, \bibinfo{person}{Yacine Jernite}, \bibinfo{person}{Julien Plu}, \bibinfo{person}{Canwen Xu}, \bibinfo{person}{Teven Le~Scao}, \bibinfo{person}{Sylvain Gugger}, \bibinfo{person}{Mariama Drame}, \bibinfo{person}{Quentin Lhoest}, {and} \bibinfo{person}{Alexander Rush}.} \bibinfo{year}{2020}\natexlab{}.
\newblock \showarticletitle{Transformers: State-of-the-Art Natural Language Processing}. In \bibinfo{booktitle}{\emph{Proceedings of the 2020 Conference on Empirical Methods in Natural Language Processing: System Demonstrations}}, \bibfield{editor}{\bibinfo{person}{Qun Liu} {and} \bibinfo{person}{David Schlangen}} (Eds.). \bibinfo{publisher}{Association for Computational Linguistics}, \bibinfo{address}{Online}, \bibinfo{pages}{38--45}.
\newblock
\urldef\tempurl%
\url{https://doi.org/10.18653/v1/2020.emnlp-demos.6}
\showDOI{\tempurl}


\bibitem[Zhu et~al\mbox{.}(2022)]%
        {Zhu_2022}
\bibfield{author}{\bibinfo{person}{Yiming Zhu}, \bibinfo{person}{E. Haq}, \bibinfo{person}{Lik-Hang Lee}, \bibinfo{person}{Gareth Tyson}, {and} \bibinfo{person}{Pan Hui}.} \bibinfo{year}{2022}\natexlab{}.
\newblock \showarticletitle{A Reddit Dataset for the Russo-Ukrainian Conflict in 2022}.
\newblock \bibinfo{journal}{\emph{preprint arXiv:2206.05107}} (\bibinfo{year}{2022}).
\newblock
\urldef\tempurl%
\url{https://doi.org/10.48550/arxiv.2206.05107}
\showDOI{\tempurl}


\end{thebibliography}

\end{document}